\documentstyle[preprint]{jpsj}

\def\runtitle{
Anomalous Magnetism of UPt$_{3}$
}
\def\runauthor{Yukihiro {\sc Okuno} and Kazumasa {\sc Miyake}}

\title
{
Anomalous Magnetism of UPt$_{3}$: \\
The Possibility of  Oscillating Antiferromagnetic Moment
}

\author
{
Yukihiro {\sc Okuno}${^{1,2}}$
\footnote{Present address : Yukawa Institute for Fundamental Physics,
Kyoto University, Kyoto 606-8502}
and Kazumasa {\sc Miyake}${^2}$
}

\inst
{
${^1}$Department of Physics, Faculty of Science,
Kyoto University, Kyoto 606-8502 \\
${^2}$Department of Physical Science, Graduate School  of Engineering Science,
Osaka University, Toyonaka 560-8531
}

\recdate
{ May 13, 1998
}
\abst
{ UPt${}_{3}$ exhibits a paradoxical behavior in its antiferromagnetism.
Namely, the  neutron scattering experiments surely detect 
the antifferomagnetic ordering 
below the transition temperature ${\rm T}_{N}\sim 5K$, while  
no trace of the transition is seen in the NMR experiments and in  
the specific heat at ${\rm T}_{N}$. 
For resolution of this paradox, 
we propose  
the  ``SDW state'' which oscillates with a finite frequency 
larger than those of the NMR and smaller than those of the 
neutron scattering. For 
demonstrating such 
possibility we seek the divergent point of susceptibility, and 
offer a preliminary calculation of the frequency dependence of 
``magnetization''. 
}
\kword
{UPt${}_{3}$, 
oscillating SDW state,  mode-mode coupling theory 
of nested spin fluctuations  
}
\begin{document}
\sloppy
\maketitle
\quad  The heavy fermion compound 
UPt${}_{3}$ has paradoxical magnetic properties.
Neutron scattering experiments show the onset of 
antiferromagnetic order at T${}_{N}$=5K~\cite{rf:Aeppli}
with the presence of 
ordered moment equal to 0.02$\mu_{B}$ at low temperatures. 
On the other hand, NMR experiments cannot detect any trace
of the ordered moment and the specific heat has 
no signal of the transition at all.~\cite{rf:Tou}
These experimental facts 
are considered to be well established. These apparently 
contradictory results 
may be understood naturally if the  
``antiferromagnetism'' detected by neutron scattering 
is not a true order but fluctuates with some finite frequencies 
smaller than $\sim 10^{11}$/s, characteristic of 
the neutron scattering, but greater than $\sim 10^{7}$/s, 
which are those of the NMR measurements. 
However the correlation length, $\xi$, is about $300{\rm \AA}$
which is so long that it is nearly equal to the resolution 
limit of the neutron scattering experiments. In 
addition, it can be observed over a wide temperature range 
from the ``transition temperature'' 5K down to 
zero temperature~\cite{rf:Koike}, this fact indicates that 
the spin fluctuation does not obey the ordinary scaling law, 
$\xi \propto |(T-T_{c})/T_{c}|^{-\nu}$.
We consider  that this  apparent paradox indicates   
 a novel phenomenon and   
we propose here as its solution  
that the ``SDW state''
oscillates with a finite frequency. \\
\indent First we explore the possibility that
the divergence of staggered susceptibility 
occurs at a finite frequency 
rather than at zero frequency. If it is possible,
it strongly suggests the existence of the ``order'' 
which  may oscillate with some finite frequency. 
The ``order'' we mean here is a kind 
of spin fluctuation which can be 
regarded as an antiferromagnetic 
state if observed for a very short time interval 
like the neutron scattering investigate, in other words, 
we may call it a ``dynamical order''.
We search the origin of such an exotic state for the nesting
property of the Fermi surface 
which causes strong frequency dependence to 
the mode-mode coupling coefficient 
The nesting property of the Fermi surface 
of UPt${}_{3}$ is partly supported  by the band 
structure calculations and the dHvA effect.~\cite{rf:Kimura}
We calculate the frequency dependence of ``transition temperature'' 
using  the  mode-mode coupling scheme  
with a mode coupling coefficient which 
has a peculiar frequency dependence. \\
\indent On the basis of the itinerant-localized duality
 model,~\cite{rf:Kuramoto,rf:Miyake} the dynamical
spin susceptibility  $\chi (Q+q,{\rm i}\omega_{m})$ for
the coherent spin fluctuations is given as, $
\chi (Q+q,{\rm i}\omega_{m})
 = \chi_{s}(Q+q,{\rm i}\omega_{m})
+ \chi_{f}(Q+q,{\rm i}\omega_{m})$, 
where $\chi_{s}(Q+q,{\rm i}\omega_{m})$ is due to the localized 
components of spins,
  $\chi_{f}(Q+q,{\rm i}\omega_{m})$ 
to the itinerant fermions, and ${\rm Q}$ is the commensurate
antiferromagnetic wave vector. \\
\indent The localized spin part $\chi_{s}(Q+q,{\rm i}\omega_{m})$ is
obtained using the mode-mode coupling theory as
\begin{equation}
\chi_{s}^{-1}(Q+q,{\rm i}\omega_{m}) =
 \chi_{s0}^{-1}(Q+q,{\rm i}\omega_{m})
-\Pi(Q+q,{\rm i}\omega_{m}), \label{chis}
\end{equation}
where the dynamical susceptibility for the localized spin
$\chi_{s0}(Q+q,{\rm i}\omega_{m})$ without the coupling
to the itinerant fermion is given as 
$\chi_{s0}^{-1}(Q+q,{\rm i}\omega_{m})=
 \chi_{0}^{-1}({\rm i}\omega_{m})-J(Q+q)$, 
where  $\chi_{0}({\rm i}\omega_{m})$ is the local susceptibility
and $J(Q+q)$ the RKKY-type interaction
among the localized component of spins. \\
\indent The fermion polarization $\Pi$ is given as  
$\Pi =\Pi_{0}+\Pi_{1}$ where $\Pi_{0}$ is 
 the bare electron-hole polarization part and the $\Pi_{1}$ is the 
mode-mode coupling term, which are given as \\
\begin{subeqnarray}
&&\Pi_{0}(Q+q,{\rm i}\omega_{m})=
-g^{2}T\sum_{n,p}G(p,{\rm i}\epsilon_{n})G(p+Q+q,
{\rm i}\epsilon_{n}+{\rm i}\omega_{m}),  \nonumber \\
&&\hspace{2.0cm} =-g^{2}T\sum_{n,p}\frac{z^{2}}
{({\rm i}\epsilon_{n}-\xi_{p})
({\rm i}\epsilon_{n}+{\rm i}\omega_{m}-\xi_{p+Q+q})}  \\
&&\Pi_{1}(Q+q,{\rm i}\omega_{m})=-g^{4}T\sum_{m',q'}G_{4}
\chi_{s}(Q+q',{\rm i}\Omega_{m'}),  \label{Pi} 
\end{subeqnarray}
where $G(p,{\rm i}\epsilon_{n})$ is the Green function for the
itinerant fermion, $z$ is the renormalization amplitude, 
and $g$ is the coupling constant between
localized spin fluctuations and fermions and is related to the
fermion coherent temperature $zg\sim T_{0}$~\cite{rf:Kuramoto}.
\begin{figure}
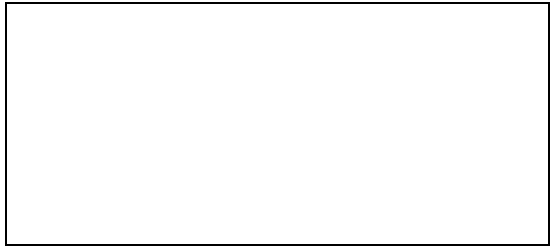

\figureheight{3.0cm}
\caption{The Feynmann diagram for the 
mode-mode coupling coefficient, $G_{4}$. The solid line represents the 
fermion propagator and the wavy line 
the spin fluctuation propagator. The dot represents the spin-fermion coupling
coefficient. } 
\label{fig:1}
\end{figure}
The mode-mode coupling coefficient ($G_{4}$) is given as 
$G_{4}=6G^{s}_{4} - G^{v}_{4}$ 
where 
\begin{subeqnarray}
\lefteqn{G^{s}_{4}=T\sum_{n}\sum_{p}G^{2}(p,{\rm i}\epsilon_{n})
G(p+Q+q,{\rm i}\epsilon_{n}+{\rm i}\omega_{m})}\\ 
&&\times G(p+Q+q',{\rm i}\epsilon_{n}+{\rm i}\Omega_{m'}), \nonumber \\
\lefteqn{G^{v}_{4}=T\sum_{n}\sum_{p}G(p,{\rm i}\epsilon_{n})
G(p+Q+q,{\rm i}\epsilon_{n}+{\rm i}\omega_{m})}\\
&&\times G(p+q-q',{\rm i}\epsilon_{n}+{\rm i}\Omega_{m'})
G(p+q'-Q,{\rm i}\epsilon_{n}-{\rm i}\Omega_{m'}). \nonumber 
\end{subeqnarray}
\indent The polarization $\Pi_{0}$, given by eq.(\ref{Pi}a), 
 is expanded for small $|q|$ and
small $\omega$ as, $
\Pi_{0}(Q+q,\omega) \simeq N_{F}^{-1}[1-A_{f}q^{2}+{\rm i}C\omega],$ 
where we assume 
 $\Pi(Q,0)$ to be $N_{F}^{-1}$ with $N_{F}$ being the
effective density of states. 
The coefficients $A_{f}$ and $C$ are given as~\cite{rf:Miyake}
$A_{f} = a_{f}v_{F}^{2}/[{\rm max}(T,h)]^{2}$ and 
$C = c_{f}/{\rm max}(T,h),$ respectively, 
where $a_{f}$ and $c_{f}$ are numerical constants, $v_{F}$
is the renormalized Fermi velocity, and h is the parameter 
measuring the deviation from the perfect nesting condition.
The nesting property of the systems is
reflected in the strong temperature dependence of the coefficient
$A_{f}$ and $C$ in contrast to the systems without nesting.
So the retarded counterpart of $\chi_{s}(Q+q,{\rm i}\omega_{m})$ is
expressed as
\begin{equation}
\chi_{s}(Q+q,\omega) = \frac{N_{f}}
{A[\kappa(\omega)^{2}+q^{2}]-{\rm i}C\omega},
\end{equation}
where  $\kappa$ is the inverse of the correlation length,  
$A = A_{f}+a_{s}Ja^{2}N_{f}$   
where  $a_{s}$  are numerical constants due to the localized spins, 
$J$ is the energy scale of the RKKY-type interaction and  $a$ is the
lattice constant~\cite{rf:Miyake}.\\
\indent Due to the nesting property, $G_{4}$ in eq.(\ref{Pi}b)
 not only has a temperature
dependence but also has a strong frequency dependence. 
Here we calculate the mode-mode coupling coefficient 
assuming that there exists a 
`dominant frequency' $i\omega_{m}$ as well as a wave vector Q.
The mode-mode coupling coefficient, which is shown in Fig. 1, 
is obtained after performing frequency summations as,
\begin{eqnarray}
&&G_{4}
=z^{4}g^{4}\sum_{p}({\rm th}\frac{\xi_{p}}{2{\rm T}}
-{\rm th}\frac{\xi_{p+Q}}{2{\rm T}})\frac{1}
{({\rm i}\omega_{m}+\xi_{p}-\xi_{p+Q})^{3}}  \\
&&-\frac{z^{4}g^{4}}{4{\rm T}}\sum_{p}
({\rm cosh}^{-2}\frac{\xi_{p}}{2{\rm T}}+
{\rm cosh}^{-2}\frac{\xi_{p+Q}}{2{\rm T}})
 \frac{1}{({\rm i}\omega_{m}+\xi_{p}-\xi_{p+Q})^{2}} \nonumber \label{Gmode}
\end{eqnarray}
When the perfect nesting condition $\xi_{p+Q} = -\xi_{p}$ is fulfilled, 
we can  see that $G_{4}$, given by eq.(\ref{Gmode}),  
has  asymptotic dependence $1/\omega_{m}^{2}$ as 
$|\omega_{m}|\rightarrow \infty$. In Fig. 2
we also show the calculated result of frequency dependence of 
(\ref{Gmode}) with the model dispersion 
$\xi_{p}=-2t({\rm cos} {\it k_{x}}a + {\rm cos} {\it k_{y}}a)-\mu$ on 
the square lattice for various fillings.   
We can see the strong $\omega_{m}$ 
dependence which drastically 
decreases the mode-mode coupling coefficient with 
asymptotically proportional to  $1/\omega_{m}^{2}$. 
This marked decrease in mode-mode 
coupling coefficient reflects the 
nesting property and is important for our discussion.
\begin{figure}
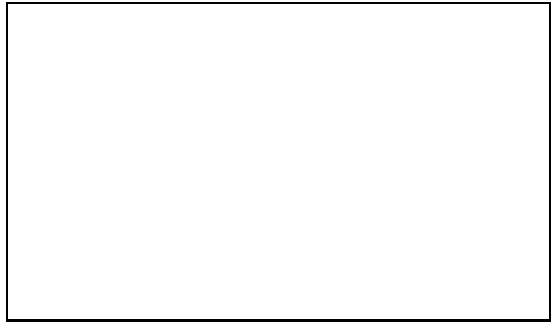

\figureheight{4cm}
\caption{Frequency dependence of the mode-mode 
coupling coefficient with the model dispersion 
$\xi_{p}=-2t(\cos k_{x}+\cos k_{y})$ for various filling n.
The bold line represents the slope of $\omega^{-2}$ for a guide to an eye.}
\label{fig:2}
\end{figure}
Combining the asymptotic dependence of temperature~\cite{rf:Miyake},         
we can take the interpolating dependence of $G_{4}$ on
${\rm T}$, ${\rm h}$ and $\omega_{m}$, for simplicity,
\begin{equation}
G_{4} \simeq g_{f}z^{4}g^{4}N_{F}
\left(\frac{\tau^{2}}{\omega_{m}^{4}+\tau^{4}}+
\frac{\omega_{m}^{2}}{\omega_{m}^{4}+\Gamma^{2}}\right) \label{mode2}
\end{equation}
 where  $\tau \equiv \sqrt{T^{2}+{\rm h}^{2}}$ and  
$g_{f}$ is a numerical constant from fermion
contribution and is given as $g_{f}=35\zeta (3)/8\pi^{2}$ for the 3D 
spherical band and $\Gamma$ is the parameter
which is introduced so as to reproduce the correct asymptotic form. 
We choose this interpolating
form so as to reproduce the correct asymptotic form concerning 
dependence on $\omega_{m}$ and $T$, and 
not to lead to unrealistic singularities when 
 $G_{4}({\rm i}\omega_{m})$ is analytically continued onto  
  the real axis, {\it i.e.}, 
${\rm i}\omega_{m}\rightarrow \omega+{\rm i}\delta$. 
Considering that $\chi (Q,{\rm i}\omega_{m})$ takes real 
values and is as 
even function of $\omega_{m}$, we may be able to 
use  eq.(\ref{mode2}) as an appropriate form.
With this frequency dependent
mode-mode coupling coefficient, we follow the conventional
mode-mode coupling theory~\cite{rf:Moriya}. 
The self-consistent equation is derived 
from (\ref{chis}) (\ref{Pi}a) and (\ref{Pi}b) as
\begin{eqnarray}
\hspace{-1cm}&&y({\rm i}\omega_{m})\equiv
\frac{N_{F}}{Aq_{B}^{2}}(\chi^{-1}_{s0}(Q,0)-\Pi_{0}(Q,0)) \nonumber \\
&&+\frac{N_{F}}{Aq_{B}^{2}}G_{4}({\rm i}\omega_{m})
\sum_{q}{\rm T}\sum_{m'}\chi_{s}(Q+q,{\rm i}\omega_{m'}) \label{SCRIM}
\end{eqnarray}
Then we arrange it as follows,
\begin{eqnarray}
\leftline{$\displaystyle{y(i\omega)=y_{0}\frac{\tau^{2}}{1+\alpha \tau^{2}}
+Fg(\omega)\frac{\tau^{5}}{(1+\alpha \tau^{2})^{3}}} $} 
 {\label{SCRRE}} \\
\leftline{$\displaystyle{\times\int_{0}^{1}dxx^{2}
\int_{0}^{\infty}d\omega\frac{1}
{{\rm e}^{\omega /t}-1}
\frac{{\omega}}{[y(\omega)+x^{2}]^{2}+
[\frac{{\omega\tau}}{2\pi\beta(1+\alpha\tau^{2})}]^{2}}}$},
 \nonumber 
\end{eqnarray}
where 
\begin{equation} 
g(\omega_{m})
\equiv \frac{\tau^{2}}{\omega_{m}^{4}+\tau^{4}}+
\frac{\omega_{m}^{2}}{\omega_{m}^{4}+\Gamma^{2}},{\label{gomega}}
\end{equation}
and $t$, $\tau$, ${\rm h}$, and $\omega_{m}$ have been  
normalized by dividing by $N_{F}^{-1} 
\sim \epsilon_{F}$, and 
$F\equiv g_{f}N_{F}^{4}z^{4}g^{4}q_{B}^{3}/2\pi^{4}a_{f}^{2}$ 
is the constant which is proportional to the strength of mode-mode coupling.
$y_{0}\equiv (N_{F}/a_{f})[\chi_{s0}{}^{-1}(Q,0)-\Pi_{0}(Q,0)]$ 
is the measure of the distance from the magnetic phase boundary at $T=0$.
$\alpha\equiv 4a_{s}J/\pi a_{f}\epsilon_{F}$ is 
the normalized RKKY-type interaction 
 and $\beta=a_{f}/2\pi c_{f}$ is the numerical
factor. For the 3D spherical band,  
$\alpha$ and $\beta$ are given
as $\alpha=(96\pi /7\zeta (3))J/\epsilon_{F}$ and
$\beta=7\zeta(3)/8\pi^{4}$, respectively. 
$q_{B}$ is the physical cut off wave number and is 
given  as $q_{B}{}^{3}=6\pi^{2}/v_{0}$ where 
$v_{0}$ is the volume per magnetic atom.
$y(\omega)$ in the integral (\ref{SCRRE}) is  analytically
continued from $y({\rm i}{\omega_{m}})$ onto the real axis 
in complex $\omega$ plane.
The self-consistent equation for $y(i\omega)$ is written as
\begin{equation}
y(i\omega_{m}) = y_{0}\frac{\tau^{2}}{1+\alpha \tau^{2}} +
Fg(\omega_{m})\frac{\tau^{5}}{(1+\alpha \tau^{2})^{3}}
\Xi (T), \label{eta}
\end{equation}
where
\begin{eqnarray}
&& \Xi(T) \equiv \label{eta2} \\ 
&& \int_{0}^{1}dxx^{2}
\int_{0}^{\infty}d\omega\frac{1}
{{\rm e}^{\omega /t}-1}
\frac{{\omega}}{[y(\omega)+x^{2}]^{2}+
[\frac{{\omega\tau}}
{2\pi\beta(1+\alpha\tau^{2})}]^{2}}.  \nonumber 
\end{eqnarray}
\indent First we have to determine the parameter $\Xi (T)$ 
self-consistently from eqs.(\ref{eta}) and (\ref{eta2}). 
Then we determine the ``transition temperature'' ${\rm T}_{N}({\rm i}\omega)$
from the equation, $\chi_{s}(Q,{\rm i}\omega)^{-1} = 0$, {\it i.e.} 
$y({\rm i}\omega )=0$, 
 for finite frequency
${\rm i}\omega$ for which ${\rm T}_{N}({\rm i}\omega)$ 
is expected to be 
higher than  ${\rm T}_{N}(0)$.  Namely,  ${\rm T}_{N}({\rm i}\omega)$ 
is determined by the condition 
\begin{equation}
y_{0}\tau +\frac{1}{2\pi\beta}
\omega+Fg(\omega)
\frac{\tau^{4}}{(1+\alpha\tau^{2})^{2}}\Xi (T) = 0, \label{Tndet}
\end{equation}
where the frequency ${\rm i}\omega$ is analytically continued 
on the imaginary axis from the Matsubara frequency points 
${\rm i}\omega_{m}$'s. \\
\indent The numerical solution of eq.(\ref{Tndet}) 
is shown in Fig. 3.
We can see that the highest 
``transition temperature'' ${\rm T}_{N}$ is
obtained at the finite frequency $i\omega_{0}$.
The abrupt disappearance of the 
transition temperature in the  high frequency 
region is due to the $\omega$-dependent structure 
of eq.(\ref{Tndet}), 
which may not be valid in this region. 
A realistic physical 
situation may be  that the ``transition temperature''  
once reaches its 
maximum at some frequency, ${\rm i}\omega_{0}$, and then gradually 
approaches zero with increasing  frequency ${\rm i}\omega$.
Hereafter we write the  highest ``transition temperature'' as 
${\rm T}_{N}(i\omega_{0})$. \\
\indent The reason for such frequency dependence
of ${\rm T}_{N}$ is easily understood by inspecting the structure of
eq.(\ref{Tndet}). 
The mode-mode coupling term of eq.(\ref{Tndet}),
which plays the role of reducing 
the transition temperature ${\rm T}_{N}$,
is suppressed by 
the factor $g(\omega)$, given by eq.(\ref{gomega}), 
with increasing frequency $\omega$ so that the effect of increasing
${\rm T}_{N}$  overwhelms the damping term arising from the 
electron-hole pair excitations, the second term of 
eq.(\ref{Tndet}), in some region of frequency. \\
\begin{figure}
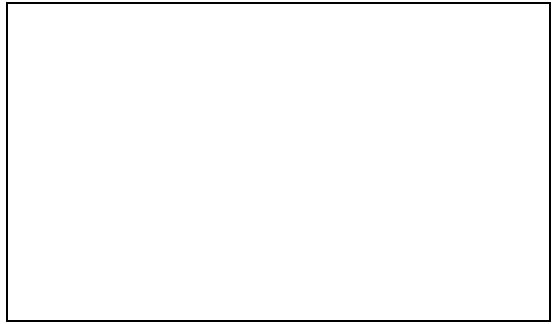

\figureheight{4cm}
\caption{
``Transition temperature'' ${\rm T}_{N}({\rm i}\omega)$ 
as a solution of eq.(\protect{\ref{Tndet}}). 
} 
\label{fig:3}
\end{figure}
\indent In Fig. 3 we also show the result with large damping 
(small value of $\beta$ in eq.(\ref{Tndet}) ),
in which case the transition temperature 
takes the highest value at 
zero frequency like the ordinary SDW case.
Thus we obtain a possible structure 
of the magnetic susceptibility that has 
divergence at some finite frequency under certain conditions.
We recognize this 
as an instability due to the emergence of 
the dynamical component of the SDW moment. So the state we consider below 
is different from 
the usual paramagnetic phase and in this context it can be 
said that it is a phase transition but the order 
in this state cannot 
be defined for a long time interval.\\ 
\indent By a simple argument we can expect that 
the ``SDW moment'' ${\rm M}_{Q}$ 
has its peak at some finite 
value of real frequency $\omega$.
If the mode-mode  
coupling coefficient given by eq.(\ref{mode2}),  
which corresponds to the fourth-order term in GL expansion,  
has its peak at $\omega=0$, the existence of 
zero-frequency component of ``order parameter'' involved 
more energy 
than the other components of frequency. 
In order  to confirm the above prediction 
we calculate the SDW ``moment'' on the real $\omega$-axis. \\
\indent For simplicity,
we retain only one frequency component of the ``order parameter'', 
${\rm M}_{Q}(\omega)$,  and do not take into account the  
 coupling between different frequency 
components of ${\rm M}_{Q}(\omega)$ as the first approximation.
In the presence of ${\rm M}_{Q}(\omega)$, 
the self-consistent
mode-mode coupling equation 
is given as
\begin{subeqnarray}
y_{\perp }&=&y_{0}\frac{\tau^{2}}{1+\alpha\tau^{2}}
+\frac{F}{5}g(\omega)
\frac{\tau^{5}}{(1+\alpha\tau^{2})^{3}}
(3\Xi_{\perp }+2\Xi_{\parallel }) \nonumber \\
&+&\frac{\tau^{2}}{1+\alpha\tau^{2}}g(\omega ){\rm M}_{Q}(\omega )^{2} \\
y_{\parallel }&=&y_{\perp }+{\rm M}_{Q}(\omega )
\frac{\partial y_{\perp }}{\partial {\rm M}_{Q}(\omega )}, \label{chiM}
\end{subeqnarray}
where the subscripts $\perp $ and $\parallel $ denote
 the transverse and the longitudinal
components of susceptibilities, respectively, 
and $y_{\perp }=N_{F}/Aq_{B}^{2}\chi_{\perp }(Q,\omega)$. 
We should note that the damping term, linear in ${\rm i}\omega$ 
 vanishes when the wave vector 
of $\chi_{\perp }(q,\omega)$ is just the antiferromagnetic 
wave vector Q in the presence of ${\rm M}_{Q}(\omega)$.
The $\Xi_{\nu}(T), \nu =\perp $ or $\parallel $
is given as,
\begin{eqnarray}
&&\Xi_{\nu}(T) = \\  
&& \int_{0}^{1}dxx^{2}
\int_{0}^{\infty}d\omega\frac{1}
{{\rm e}^{\omega /t}-1}
\frac{{\omega}}{(y_{\nu}(\omega)+x^{2})^{2}+
[\frac{{\omega\tau}}
{2\pi\beta(1+\alpha\tau^{2})}]^{2}}. \label{etaM} \nonumber
\end{eqnarray}
\indent The magnetization ${\rm M}_{Q}(\omega)$  
 is determined from the equation,  
$\chi_{\perp }(Q,\omega,{\rm M}_{Q}(\omega))^{-1}=0$,
where the subscript $\perp $  denotes the transverse
component of susceptibility, 
and this condition is rearranged as,
\begin{eqnarray}
&&y_{\perp }(Q,\omega)=y_{0}\tau + \frac{D}{1+\alpha\tau^{2}}\omega^{2}
\label{detM} \\
&&+\frac{F}{5}g(\omega)\frac{\tau^{4}}
{(1+\alpha\tau^{2})^{2}}(3\Xi_{\perp }+2\Xi_{\parallel })  
+\tau g(\omega) {\rm M}_{Q}(\omega )^{2}=0.  \nonumber 
\end{eqnarray}
Here we have inserted  the 
$\omega^{2}$-term in eq.(\ref{detM}) which is the leading 
term in $\omega$ of the 
 susceptibility for $q=Q$  and the coefficient $D$ should  
be determined by the band structure.
We can determine  ${\rm M}_{Q}(\omega)$ by solving the coupled equations 
(\ref{chiM}) $\sim$ (\ref{detM}).
For simplicity we approximate  $\Xi_{\parallel }$ by $\Xi_{\perp }$.
This treatment is valid at 
low values of ${\rm M}_{Q}(\omega )$. 
Since we are detecting the 
frequency dependence of the ``order parameter'' 
${\rm M}_{Q}(\omega )$
near the ``transition temperature'', this 
approximation may be sufficient for
our purpose. 
The plot of ${\rm M}_{Q}(\omega )$ is shown in Fig. 4.
\begin{figure}
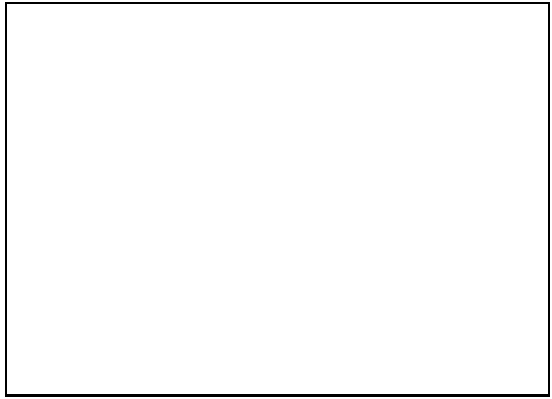

\figureheight{5cm}
\caption{
Magnitude of the ``ordered moment'' $M_{Q}(\omega)$ on real $\omega$-axis. 
}
\label{fig:4}
\end{figure}
We can see that ${\rm M}_{Q}(\omega )$ has large values at
finite frequency than at $\omega$=0.  Furthermore, 
${\rm M}_{Q}(\omega )$ 
 vanishes for low values of $\omega$ 
 (the case t=0.1 in Fig. 4). Such a situation is expected.
There exists no ${\rm M}_{Q}(0)$, the static component, 
so that it is not a true SDW state.
We should note 
that a finite value of ${\rm M}_{Q}(\omega)$ is 
only meaningful
in the time interval $\sim 2\pi/\omega $, so that the   
existence of ${\rm M}_{Q}(\omega )$ 
indicates that there exists a  
spin fluctuation mode whose ``correlation length'' is extremely  long 
and it as if there was a true  
SDW order  when observed in the time interval 
$\sim 2\pi /\omega $. The specific heat will not be changed  
by the appearance of 
${\rm M}_{Q}(\omega )$  unless 
${\rm M}_{Q}(0)$ appears because the specific 
heat jump 
in the usual phase transition mainly comes 
from the static component 
of  magnetization. 
So far, we have assumed many parameters with arbitrary values. 
 But we believe that the real behavior of 
UPt${}_{3}$ is just in the parameter 
region where this novel phenomenon  
can occur. The Plot of ${\rm M}_{Q}(\omega )$ has been  
obtained appears by assuming that there exists no coupling between   
the ``order parameter'' with different frequencies. 
The SDW ``order'' first develops  at some finite frequency.  
With decreasing 
temperature, the other frequency component, including the 
static (zero-frequency) component of the SDW state, 
should gradually develop in general. 
Indeed, the ordered component ${\rm M}_{Q}(\omega )$  
 can induce other components through  
the coupling term such as ${\rm M}_{Q}(\omega )^{3}\cdot 
{\rm M}_{Q}(-3\omega )$, 
where  
${\rm M}_{Q}(-3\omega)$ is the admixture frequency component. 
However, as long as the coupling term maintains  
its form similar to eq.(\ref{mode2}), 
the induced order ${\rm M}_{Q}(-3\omega)$ 
whose frequency is located in the frequency region 
where $T_{\rm N}(3{\rm i}\omega)=0$, results in a considerable 
energy cost 
and  hardly grows. 
When the temperature is decreased  and the `ordered' 
state develops sufficiently, 
 the mode-mode coupling term  no longer has the frequency dependence 
 like eq.(\ref{gomega}).
Then  the dominant 
frequency of the 
SDW moment may change from the value near the 
``transition temperature''. Therefore, there is a possibility that 
 the specific heat jump at T=20mK~\cite{rf:Schuberth}
may be interpreted as 
 that at this temperature the dominant frequency of 
the SDW moment has changed to the static component
({\rm i.e.}, $\omega$=0)  and it reveals 
its own state as the ordinary SDW state.~\cite{rf:Koike}\\
\indent In conclusion, existence of 
an ``oscillating SDW'' state can 
possibly be verified using a 
probe with time scale intermediate between
neutron scattering 
and NMR: say, $\mu$SR measurements.~\cite{rf:Kambe} 
A possibility of 
the ``oscillating SDW'' state can be seen in a recent 
numerical simulations for an extended Hubbard model with 
pair hopping terms.~\cite{rf:Imada} 
In the ordered state we can 
determined only the qualitative level near the 
`transition temperature' and further investigation is required 
to say details of this new state of magnetism.\\ 
\indent One of us (K.M.) has  benefited from helpful
comments of S. Kambe on the peculiarity of spin fluctuation 
properties of UPt${}_{3}$ 
and conversations with H. Harima and Y. \=Onuki 
on the Fermi surface of UPt${}_{3}$.
This study was started when one of us (K.M) visited 
J. Flouquet at CENG, whose hospitality and 
stimulating discussions are  gratefully acknowledged. 
This work is 
supported by the Grant-in-Aid for COE Research  
(10CE2004) from the Ministry of Education, Science, Sports and 
Culture.

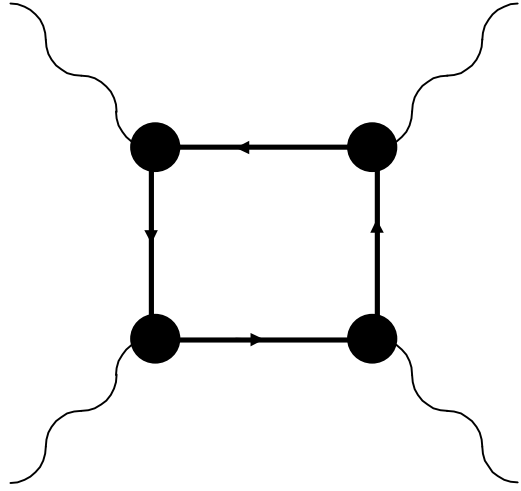
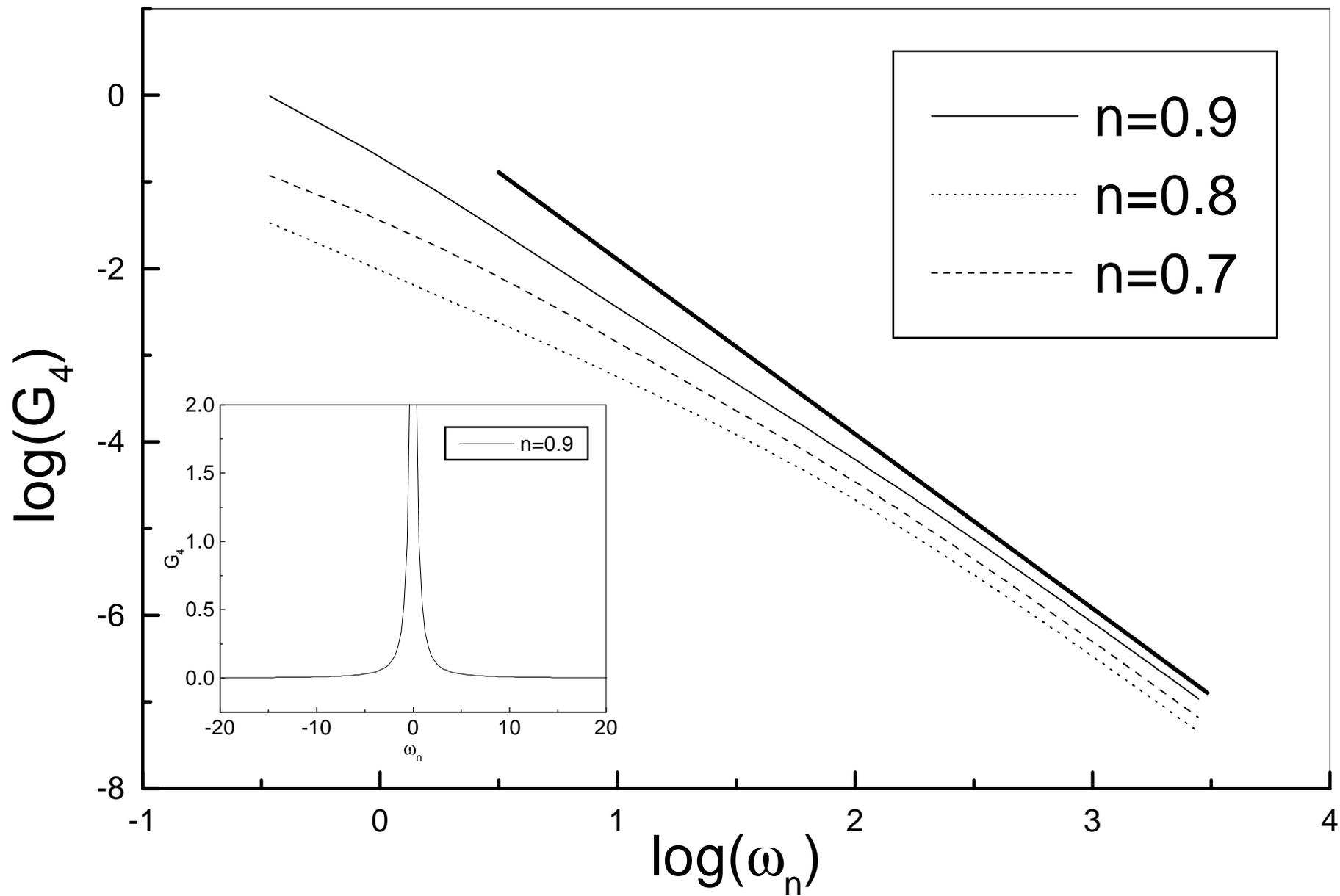
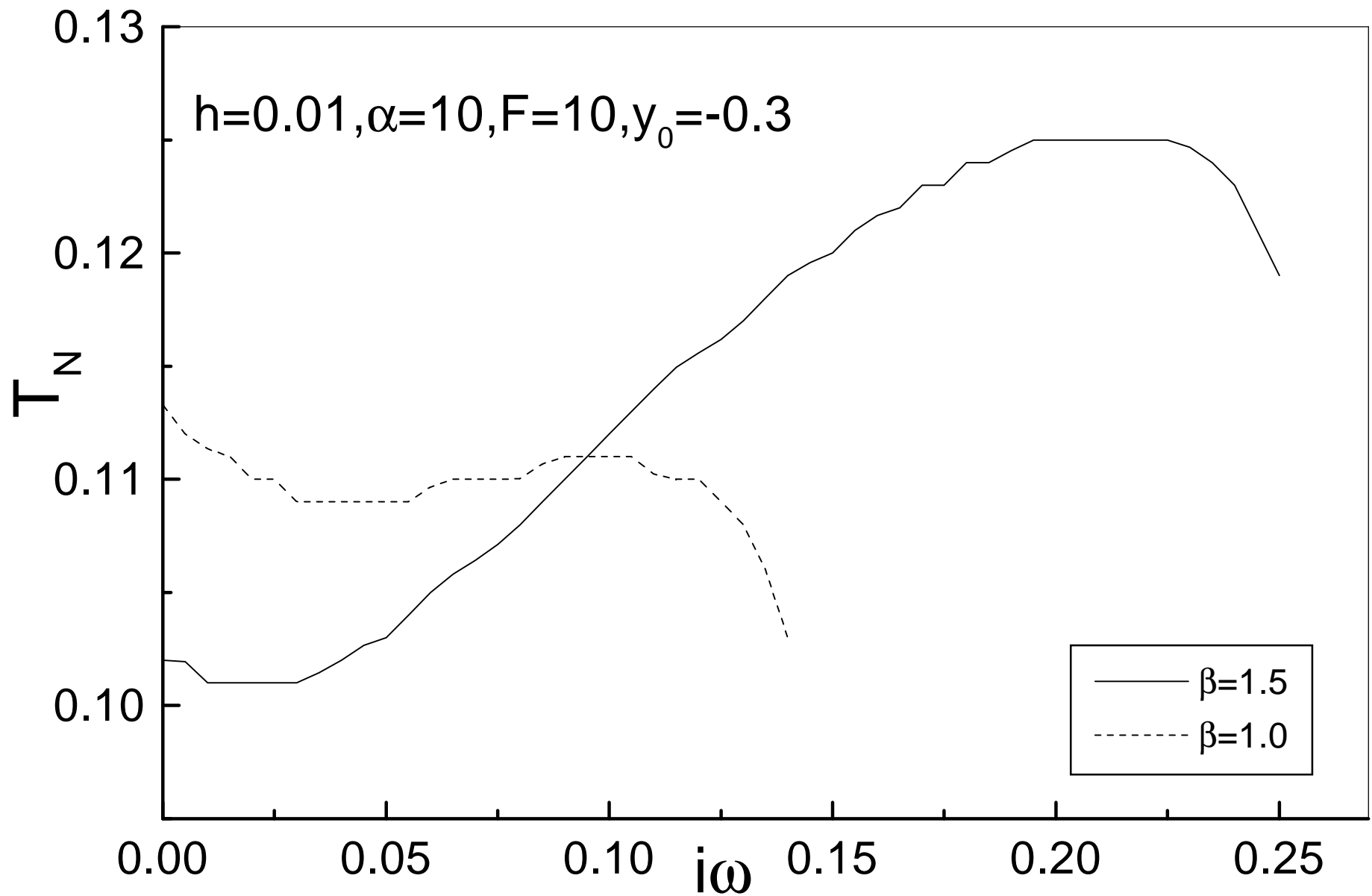
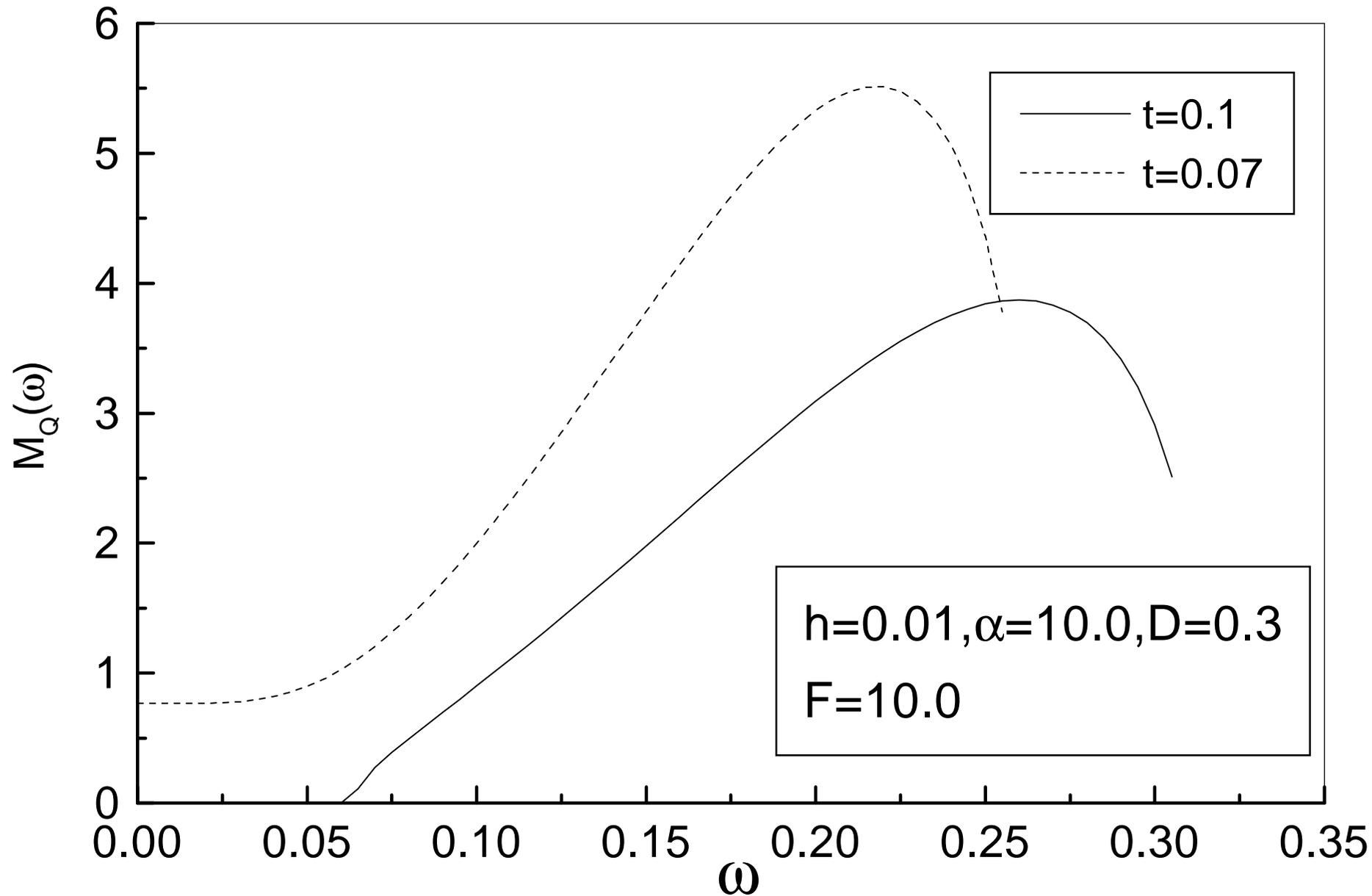
\end{document}